\documentclass[conference]{IEEEtran}
\IEEEoverridecommandlockouts
\usepackage{array,multirow,makecell}
\usepackage[T1]{fontenc}
\usepackage{multirow}
\usepackage{cite}
\usepackage[utf8x]{inputenc} 
\usepackage{longtable}
\usepackage{xcolor}
\usepackage[linesnumbered,ruled,vlined]{algorithm2e}
\usepackage{colortbl}
\usepackage{pgfplots} 

\SetCommentSty{mycommfont}
\usetikzlibrary{patterns}
\usepackage{tikz}
\usepackage{subcaption}
\usepackage{xcolor}
\usetikzlibrary{arrows,shapes,positioning,shadows,trees}
\usepackage{rotating, adjustbox}
\usepackage{pgfplots}
    \makeatletter 
\newcommand{\linebreakand}{%
  \end{@IEEEauthorhalign}
  \hfill\mbox{}\par
  \mbox{}\hfill\begin{@IEEEauthorhalign}
}
\makeatother 

\ifCLASSINFOpdf
\else
\fi
\usepackage{algorithmic}
\algsetup{linenosize=\small}
\begin{document}
%

\title{Poisoning Attacks in Federated Edge Learning for Digital Twin 6G-enabled IoTs: An Anticipatory Study}

\author{\IEEEauthorblockN{Mohamed Amine Ferrag}
\IEEEauthorblockA{
\textit{Technology Innovation Institute}\\
9639 Masdar City, Abu Dhabi, UAE \\
mohamed.ferrag@tii.ae}
\and
\IEEEauthorblockN{Burak Kantarci}
\IEEEauthorblockA{\textit{University of Ottawa} \\
Ottawa, ON, Canada\\
burak.kantarci@uottawa.ca}
\and
\IEEEauthorblockN{Lucas C. Cordeiro}
\IEEEauthorblockA{
\textit{Technology Innovation Institute}\\
9639 Masdar City, Abu Dhabi, UAE \\
lucas.cordeiro@tii.ae}
\and

\linebreakand 

\IEEEauthorblockN{Merouane Debbah}
\IEEEauthorblockA{
\textit{Technology Innovation Institute}\\
9639 Masdar City, Abu Dhabi, UAE \\
merouane.debbah@tii.ae}
\and
\IEEEauthorblockN{Kim-Kwang Raymond Choo}
\IEEEauthorblockA{\textit{University of Texas at San Antonio} \\
San Antonio, TX 78249-0631, USA\\
raymond.choo@fulbrightmail.org}
}
\maketitle

\begin{abstract}
Federated edge learning can be essential in supporting privacy-preserving, artificial intelligence (AI)-enabled activities in digital twin 6G-enabled Internet of Things (IoT) environments. However, we need to also consider the potential of attacks targeting the underlying AI systems (e.g., adversaries seek to corrupt data on the IoT devices during local updates or corrupt the model updates); hence, in this article, we propose an anticipatory study for poisoning attacks in federated edge learning for digital twin 6G-enabled IoT environments. Specifically, we study the influence of adversaries on the training and development of federated learning models in digital twin 6G-enabled IoT environments. We demonstrate that attackers can carry out poisoning attacks in two different learning settings, namely: centralized learning and federated learning, and successful attacks can severely reduce the model's accuracy. We comprehensively evaluate the attacks on a new cyber security dataset designed for IoT applications with three deep neural networks under the non-independent and identically distributed (Non-IID) data and the independent and identically distributed (IID) data. The poisoning attacks, on an attack classification problem, can lead to a decrease in accuracy from $94.93$\% to $85.98$\% with IID data and from $94.18$\% to $30.04$\% with Non-IID.
\end{abstract}


%
\IEEEpeerreviewmaketitle

\section{Introduction}

The emergence of the digital twin (DT) paradigm with 6G wireless communication networks is being viewed to transform applications and customer services through the Internet of Things (IoT) to fully autonomous and smart systems \cite{naser2022toward}. The main idea behind DT is to build a digital replication of wireless networks' physical devices and features to achieve low latency and highly reliable connectivity while providing high performance and energy efficiency in IoT networks. The architecture of digital twin 6G-enabled IoT can be organized into four layers, namely, Physical, Networking, Services, and Applications. The physical layer refers to IoT devices equipped with sensors and actuators for collecting data from the environment. The networking layer refers to DT services, telecommunication networks (6G), IEEE 802.15.4, long-range WiFi, and communication protocols to connect IoT devices with other operators and system components. The Ditto \footnote{https://www.eclipse.org/ditto/} can be used as IoT middleware, supporting an IoT abstraction level for interacting IoT solutions with physical devices via the digital twin model. The services layer provides AI, storage, computation, and services to IoT devices and Edge servers~\cite{9939166}. The applications layer refers to IoT applications, including the Internet of Vehicles, the Internet of Sensing, and the Internet of Energy...etc.



With distributed edge learning, intrusion poisoning attacks pose an even more severe challenge than the conventional machine learning environment (i.e., centralized learning). However, the threat of intrusion poisoning attacks can be challenging to overcome, as they can be difficult to identify. Nevertheless, the availability of intrusion poisoning attacks has been proven to be efficient in distributed edge learning in the following recent work:~\cite{severi2022network,zhang2022secfednids,venkatesan2021poisoning,aiken2019investigating,pawlicki2020defending,liu2021adversarial}. Zhang et al. \cite{zhang2022secfednids} proposed a defense mechanism at the model level, which is mainly based on the detection of a poisoned model. More precisely, the proposed mechanism uses the model's important parameter selection method based on the gradient to deliver the lowest dimensional efficient performances of the local model parameters downloaded.  Venkatesan et al. \cite{venkatesan2021poisoning} considered the poisoning availability attack framework, where an attacker can introduce a set of poisonous samples during training to degrade the deployed model's accuracy. Aiken et al.~\cite{aiken2019investigating} proposed an IDS system based on an anomaly. The proposed system uses a tool for adversarial testing named Hydra,  which measures the influence of the threat of adversarial evasion classifier attacks against the network intrusion detection system to reduce the detection rate of malicious network traffic.

Motivated by the facts mentioned above, in this article, we propose an anticipatory study for poisoning attacks in federated edge learning for digital twin 6G-enabled IoTs. Specifically, we demonstrate that attackers who conduct poisoning attacks in two different learning modes, namely, centralized learning and federated learning, can severely reduce the model's accuracy and the detection rate of each intrusion. We comprehensively evaluate our attacks on a new cyber security dataset designed for IoT applications, the Edge-IIoT dataset. We use different deep-learning models for cyber security intrusion detection. Furthermore, federated edge learning is evaluated under two data distribution types: IID and Non-IID data. The study demonstrates that attackers who conduct poisoning attacks in two different learning modes, namely, centralized learning and federated learning,  can lead to a decrease in accuracy from 94.93\% to 85.98\% with IID data and from 94.18\% to 30.04\% with Non-IID. 



\begin{table}
\centering
\caption{Notations used}  
\label{tab:notation}
\tiny
\setlength{\tabcolsep}{3pt}
\begin{tabular}{|c|c|} \hline
\textbf{Notation} & \textbf{Description}\\\hline
 $\eta$ & The learning rate  \\\hline
 $Epo$ & The number of local epochs  \\\hline
 $Batch$   &  The local minibatch size \\\hline
 $K_H$   &  The honest clients \\\hline
 $k_h$   &  The indexation of honest clients  \\\hline
 $K_M$   &  The malicious clients \\\hline
 $k_m$   &  The indexation of malicious clients \\\hline
 $f_{t+1}^{k_h}$ & The new local set of weights by honest client $k_h$ \\\hline
$f_{t+1}^{k_m}$ & The new local set of weights by malicious client $k_m$ \\\hline
 $C_h$   &  The fraction of honest clients*\\\hline
 $C_m$   &  The fraction of malicious clients*\\\hline
 $n_{k_h}$ &  The number of local examples for honest client $k_h$\\\hline
 $n_{k_m}$ &  The number of local examples for malicious client $k_m$\\\hline
$t$  & The number of rounds \\\hline
$S_h$    & The random set of honest clients  \\
  \hline
 $S_m$     &   The random set of malicious clients \\
  \hline
 $x$    & The weight  \\  \hline
 $\mathcal{P}_h$ & The local preprocessed dataset honest clients  \\
  \hline
   $\mathcal{P}_m$ & The local preprocessed dataset malicious clients  \\
  \hline
   $b$   &  The minibatch used for the local epoch \\
  \hline
  $ \eta	\nabla f_c(x, b)$ & The average gradient \\
  \hline
\end{tabular}\\
**The clients who compute at each round.
\end{table}

\begin{algorithm}[htp]
    \SetAlgoLined\DontPrintSemicolon
    \SetKwFunction{func}{}
    \SetKwFunction{proc}{}
      \scriptsize
    \textbf{Data}: $\eta$, $Epo$, $Batch$, $K_H$, $k_h$, $K_M$, $k_m$, $C_h$, $C_m$.\\
    \SetKwProg{myfunc}{Edge Server}{}{}
    \myfunc{EdgeFedLearn \func{$K_H$, $K_M$, $C_h$,$C_m$, $R$}:}{
    \tcc{Model initialization} 
        $f_{1} \leftarrow InitializeModel()$ \;
        \tcc{Start FEL with randomly selected clients at each round} 
        \For{$t = 1,..,R$}
        {
            $S_h$ $\leftarrow$ Subset(max($C_h\cdot K_H, 1$), $"random"$)\;
            \textbf{Parallel.}\For{$k_h$ $\in$ $S_{t}$ }
                {
                    $f_{t+1}^{k_h}$ $\leftarrow$ $ClientUpdate(f_{t}, k_h$)  \tcp*{Compute local updates of the $K_H$ honest clients using Algorithm \ref{alg:alg2}}
                }
        }
        \For{$t = 1,..,R$}
        {
            $S_m$ $\leftarrow$ Subset(max($C_m\cdot K_M, 1$), $"random"$)\;
            \textbf{Parallel.}\For{$k_m$ $\in$ $S_{t}$ }
                {
                    $f_{t+1}^{k_m}$ $\leftarrow$ $ClientUpdate(f_{t}, k_m$)  \tcp*{Compute local updates of the $K_M$ malicious clients using Algorithm \ref{alg:alg3}}
                }        }
        $f_{t+1}$ $\leftarrow$ $\sum_{k_h=1}^{K_H} \frac{n_{k_h}}{n} f_{t+1}^{k_h}$ + $\sum_{k_m=1}^{K_M} \frac{n_{k_m}}{n} f_{t+1}^{k_m}$  \tcp*{Aggregate all client updates}

        Broadcast $f_{t+1}$ the updated model to clients $K_H$ and $K_M$
        }
\caption{Poisoning Attack}
\label{alg:alg1}
\end{algorithm}
\begin{algorithm}[htp]
    \SetAlgoLined\DontPrintSemicolon
    \SetKwFunction{func}{}
    \SetKwFunction{proc}{}
      \scriptsize
    \setcounter{AlgoLine}{0}
     \textbf{Data}: $\eta$, $Epo$, $Batch$.\\
    \SetKwProg{myproc}{Honest IoT device}{}{}
    \myproc{ClientUpdate \proc{$f$, $k_h$}:}{
    \nl $\mathcal{B}$ $\leftarrow$ Split($\mathcal{P}_h$, $Batch$) \;\tcc{Split the local dataset $\mathcal{P}_h$ into $Batch$ local data batch}
    \nl \For{i = 1,..,$Epo$}
        {
            \For{$b$ $\in$ $\mathcal{B}$}
                {
                    $f$ $\leftarrow$ $f$ $-$ $ \eta	\nabla f_c(x, b)$  \tcp*{Honest client training}
                }
        }
    Return $f$ to Edge Server
}
\caption{Local updates of honest clients}
\label{alg:alg2}
\end{algorithm}

\begin{algorithm}[htp]
    \SetAlgoLined\DontPrintSemicolon
    \SetKwFunction{func}{}
    \SetKwFunction{proc}{}
      \scriptsize
    \textbf{Data}: $\eta$, $Epo$, $Batch$.\\
\SetKwProg{myproc}{Malicous IoT device}{}{}
    \myproc{ClientUpdate \proc{$f$, $k_m$}:}{
    \tcc{Client Identification algorithm based on the average loss change and backdoor attack}
    \nl $ClientIdentification()$ \;
    \tcc{Insert poison data to the local dataset}
    \nl $InsertPoisonData()$ \;
    \tcc{Assign wrong label to poison data}
    \nl $AssignWronglabel()$ \;
    \tcc{Starts the targeted client dropping attack using DDoS attacks}
    \nl $DropHonestClients()$ \;
    \tcc{Split the local dataset $\mathcal{P}_m$ into $Batch$ local data batch}
    \nl $\mathcal{B}$ $\leftarrow$ Split($\mathcal{P}_m$, $Batch$) \;
    \nl \For{i = 1,..,$Epo$}
        {
            \For{$b$ $\in$ $\mathcal{B}$}
                {
                    $f$ $\leftarrow$ $f$ $-$ $ \eta	\nabla f_c(x, b)$  \tcp*{Local malicous client training}
                }
        }
    Return $f$ to Edge Server
}
\caption{Local updates of malicious clients}
\label{alg:alg3}
\end{algorithm}

\section{The Anticipated Poisoning Attack on the Federated Edge Learning}


\subsection{Description of the Anticipated Poisoning Attack}

We systemize the poisoning threat models in federated edge learning based on three dimensions: Adversarial goal, Attack strategies, and Malicious client selection.

\subsubsection{Adversarial goal}

An attacker can focus on two types of poisoning attacks: a machine learning attack against availability (i.e., target all classes) and a machine learning attack against a particular class. Since the targeted attacks are considerably challenging to detect, we consider a Poisoning attack as a targeted attack against cyber security intrusion detection based on federated learning. Specifically, the Normal class is considered as a particular class, which is affected by the attack rate $\alpha$ $= [0\%, 40\%, 50\%, 60\%, 80\%]$. The attack rate $\alpha$ $= 60\%$ means that 60\% of Normal traffic is labeled as an attack.

\subsubsection{Attack strategies}

The attacker can conduct poisoning attacks (i.e.,  Label flipping attack, Targeted dropping attack, and Clean-label attack) in two types of learning methods: centralized learning and federated learning. In centralized learning mode, an attacker is considered to have access to the training dataset of the server using a backdoor attack. Then, it injects maliciously crafted data points containing normal IoT traffic labeled an attack class. In the federated learning mode, some malicious clients contain maliciously crafted data points, which send malicious updates to the server to change the model’s prediction. The generation of poison attacks in federated edge learning is presented in Algorithms \ref{alg:alg1}, \ref{alg:alg2}, \ref{alg:alg3}, and can be defined by the following steps. Table \ref{tab:notation} presents the notations used.

\begin{itemize}
\item Step 1: The Edge server initializes the first model $f_{1}$.

\item Step 2: The Edge server distributes the overall model $f_{t+1}$ to honest clients $K_H$ and malicious clients $K_M$.

\item Step 3: The malicious clients $K_M$ identify the best clients based on the average loss change through a backdoor attack.

\item Step 4: The malicious clients $K_M$ generate the targeted class samples and insert poison data into the local dataset.

\item Step 5: The malicious clients $K_M$ assign wrong label to poisoned data.

\item Step 6: The malicious clients $K_M$ start the targeted client dropping attack using DDoS attacks.

\item Step 7: The malicious clients $K_M$ calculate the poisoned update and it to the Edge server.
\end{itemize}

\subsubsection{Malicious clients selection}

We consider an intelligent attacker for the selectively honest client (i.e., to be malicious) to attack the federated edge learning to ensure maximum attack strength while simultaneously reducing the probability of detection. Based on the backdoor attack, this attacker can access the machine learning models located at the Edge server. The attacker has the following choices to consider: 

\begin{itemize}
    \item When to start the attack during the FL process? we found that the success of the poisoning attack is severe when the attackers control more than 51\% of the clients.
    \item What the knowledge of the attacker is? The threat model is a white box attack if the attacker has full knowledge of parameters, algorithms, data, and features. If he has only query access to the model, the threat model is a black box attack. We found that the success of the poisoning attack is severe under the white box attack. 
    \item Which honest clients to select to inject maliciously crafted data points containing normal IoT traffic labeled as an attack class? we found that the more attackers select early the best clients (i.e., who provides the best learning model), the more poisoning attack severe.
    \item How much local data to drop or inject for each client to maximize the entire global model corruption at the edge server? We found that it depends on the analysis of the local data for each client. 
    \item What is the best assignment strategy for assigning wrong labels? The assignment strategy can be implemented in various ways such as dropping, shuffling, swapping, and sliding. After several experiments, we found that the assignment strategy of the wrong label based on swapping affects more the distributed edge learning compared to others strategies.
\end{itemize}

\begin{figure*}
\centering
    \includegraphics[width=0.7\linewidth]{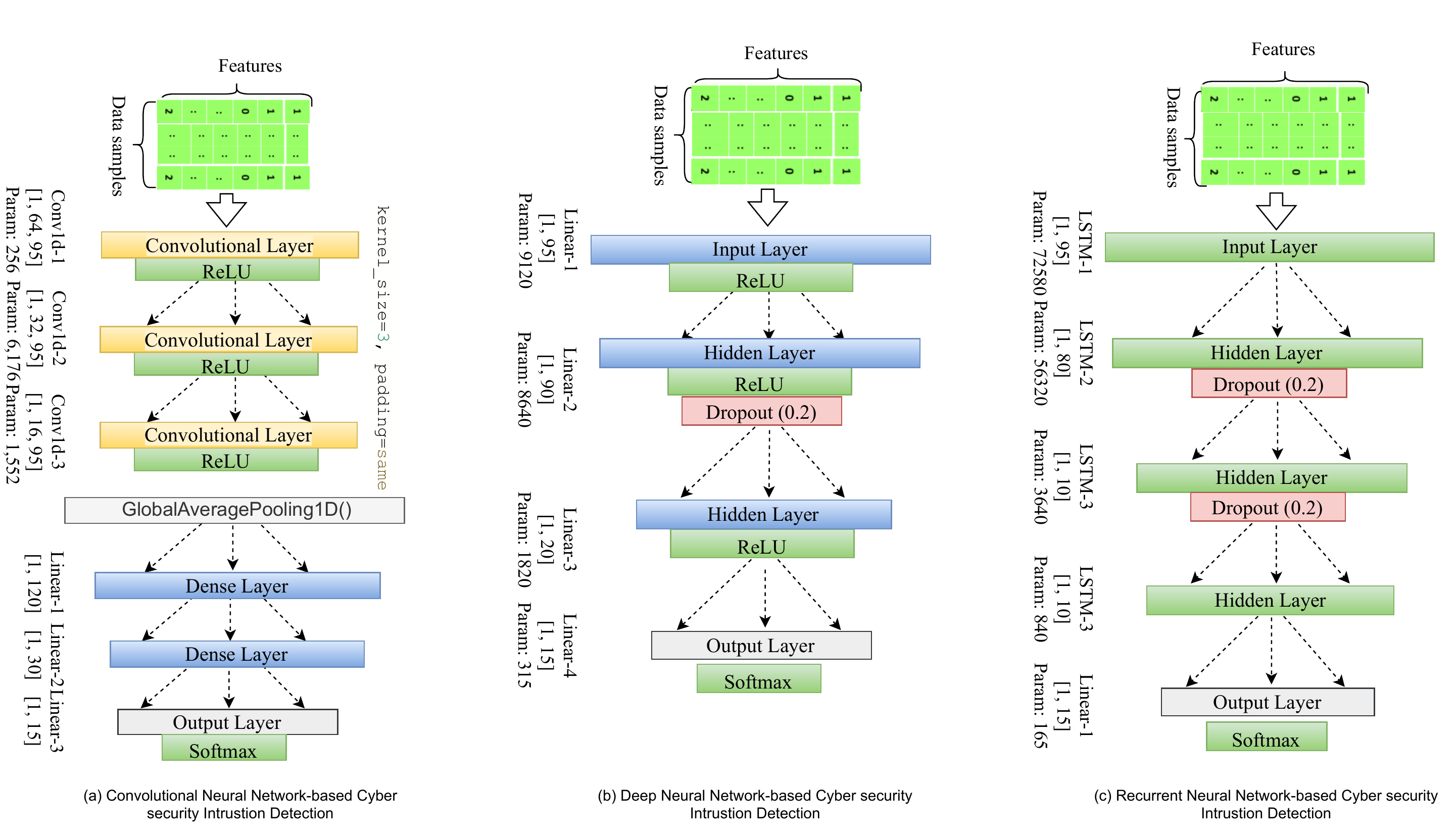}
    \caption{Structure of three deep neural models adopted by intrusion detection.}
    \label{fig:fig11}
\end{figure*}

We use three different deep-learning models for intrusion detection, namely, DNN, RNN, and CNN. The architectures of three deep neural models adopted by the intrusion detection are illustrated in Fig. \ref{fig:fig11}.

\begin{table}[h!]

\setlength{\tabcolsep}{2.5pt}
\renewcommand{\arraystretch}{1}
\caption{Settings for experimental evaluation.}
\centering
\tiny
\label{tab:tab12}
\begin{tabular}{|c|c|c|}
\hline
\textbf{}                    & \textbf{Parameter}            & \textbf{Value}                                                                                    \\ \hline
\multirow{2}{*}{Centralized} & Batch size                    & 800                                                                                               \\ \cline{2-3} 
                             & Total epochs                  & 25                                                                                                \\ \hline
\multirow{7}{*}{Federated}   & Local epochs                  & 3                                                                                                 \\ \cline{2-3} 
                             & Global epochs                 & 10                                                                                                \\ \cline{2-3} 
                             & Batch size                    & 100                                                                                               \\ \cline{2-3} 
                             & Rounds                        & 10                                                                                                \\ \cline{2-3} 
                                & Total number of IoT clients                        & 100                                                                                                \\ \cline{2-3} 
                             & Honest IoT clients used in federated updates               & 3, 7, 10, 20                                                                                          \\ \cline{2-3} 
                             & Malicious IoT clients used in federated updates            & 3, 7, 15                                                                                             \\ \cline{2-3} 
                             & Data distribution             & IID, Non-IID                                                                                      \\ \hline
\multirow{16}{*}{*}          & DL classifiers                & CNN, RNN, DNN                                                                                     \\ \cline{2-3} 
                             & Classification tasks          & \begin{tabular}[c]{@{}c@{}}Binary classification and \\ multi-class   classification\end{tabular} \\ \cline{2-3} 
                             & Number of classes             & 15                                                                                                \\ \cline{2-3} 
                             & Learning rate                 & 0.1, 0.01, 0.001                                                                                  \\
                             \cline{2-3} 
                             & Encoding function                 & One-Hot-Encoding
                                                                                  \\
                             \cline{2-3} 
                             & Kernel regularizer            & L2                                                                                                \\ \cline{2-3} 
                             & Kernel initializer            & Random uniform                                                                                    \\ \cline{2-3} 
                             & Regularization technique      & Dropout(0.2)                                                                                      \\ \cline{2-3} 
                             & Activation function           & ReLU                                                                                              \\ \cline{2-3} 
                             & Classification function       & Softmax                                                                                           \\ \cline{2-3} 
                             & Oversample the minority class & SMOTE                                                                                             \\ \cline{2-3} 
                             & Optimizer                     & Adam                                                                                              \\ \cline{2-3} 
                             & Multi-class loss function     & Categorical   crossentropy                                                                        \\ \cline{2-3} 
                             & Binary loss function          & Binary crossentropy                                                                               \\ \cline{2-3} 
                             & Metrics                       & \begin{tabular}[c]{@{}c@{}}Accuracy, Precision, \\ Recall, Confusion \\ matrix, F1-score,\\ Poisoning attack rate\end{tabular}        \\ \cline{2-3} 
                             & The threat model              & \begin{tabular}[c]{@{}c@{}}Label flipping attack,\\ Targeted dropping attack,\\ and Clean-label attack\end{tabular}        \\ \cline{2-3} 
                             & Attack rate                   & 0\%, 40\%, 50\%, 60\%                                                                             \\ \hline
\end{tabular}

\end{table}


\begin{table}[h!]

\setlength{\tabcolsep}{2.5pt}
\renewcommand{\arraystretch}{1}
\caption{Classification report for multi-class deep learning approaches (Centralized model performance).}
\centering
\label{tab:tab8}
\tiny
\begin{tabular}{|cc|ccc|ccc|ccc|}
\hline
\multicolumn{2}{|c|}{\textbf{}}                                                        & \multicolumn{3}{c|}{\textbf{Precision}}                                              & \multicolumn{3}{c|}{\textbf{Recall}}                                                 & \multicolumn{3}{c|}{\textbf{ha-score}}                                               \\ \hline
\multicolumn{1}{|c|}{\textbf{Poisoning Attack}}                & \textbf{Class}         & \multicolumn{1}{c|}{\textbf{DNN}} & \multicolumn{1}{c|}{\textbf{RNN}} & \textbf{CNN} & \multicolumn{1}{c|}{\textbf{DNN}} & \multicolumn{1}{c|}{\textbf{RNN}} & \textbf{CNN} & \multicolumn{1}{c|}{\textbf{DNN}} & \multicolumn{1}{c|}{\textbf{RNN}} & \textbf{CNN} \\ \hline
\multicolumn{1}{|c|}{\multirow{15}{*}{\begin{tabular}[c]{@{}c@{}}No attack\\   \end{tabular}}}             & Backdoor               & \multicolumn{1}{c|}{72\%}         & \multicolumn{1}{c|}{77\%}         & 72\%         & \multicolumn{1}{c|}{93\%}         & \multicolumn{1}{c|}{92\%}         & 93\%         & \multicolumn{1}{c|}{81\%}         & \multicolumn{1}{c|}{84\%}         & 81\%         \\ \cline{2-11} 
\multicolumn{1}{|c|}{}                                        & DDoS\_HTTP             & \multicolumn{1}{c|}{70\%}         & \multicolumn{1}{c|}{73\%}         & 70\%         & \multicolumn{1}{c|}{98\%}         & \multicolumn{1}{c|}{95\%}         & 99\%         & \multicolumn{1}{c|}{82\%}         & \multicolumn{1}{c|}{83\%}         & 82\%         \\ \cline{2-11} 
\multicolumn{1}{|c|}{}                                        & DDoS\_ICMP             & \multicolumn{1}{c|}{98\%}         & \multicolumn{1}{c|}{100\%}        & 99\%         & \multicolumn{1}{c|}{98\%}         & \multicolumn{1}{c|}{99\%}         & 99\%         & \multicolumn{1}{c|}{98\%}         & \multicolumn{1}{c|}{100\%}        & 99\%         \\ \cline{2-11} 
\multicolumn{1}{|c|}{}                                        & DDoS\_TCP              & \multicolumn{1}{c|}{68\%}         & \multicolumn{1}{c|}{71\%}         & 69\%         & \multicolumn{1}{c|}{100\%}        & \multicolumn{1}{c|}{100\%}        & 100\%        & \multicolumn{1}{c|}{81\%}         & \multicolumn{1}{c|}{83\%}         & 82\%         \\ \cline{2-11} 
\multicolumn{1}{|c|}{}                                        & DDoS\_UDP              & \multicolumn{1}{c|}{99\%}         & \multicolumn{1}{c|}{100\%}        & 99\%         & \multicolumn{1}{c|}{99\%}         & \multicolumn{1}{c|}{100\%}        & 100\%        & \multicolumn{1}{c|}{99\%}         & \multicolumn{1}{c|}{100\%}        & 99\%         \\ \cline{2-11} 
\multicolumn{1}{|c|}{}                                        & Fingerprinting         & \multicolumn{1}{c|}{0\%}          & \multicolumn{1}{c|}{27\%}         & 0\%          & \multicolumn{1}{c|}{0\%}          & \multicolumn{1}{c|}{84\%}         & 0\%          & \multicolumn{1}{c|}{0\%}          & \multicolumn{1}{c|}{41\%}         & 0\%          \\ \cline{2-11} 
\multicolumn{1}{|c|}{}                                        & MITM                   & \multicolumn{1}{c|}{100\%}        & \multicolumn{1}{c|}{100\%}        & 100\%        & \multicolumn{1}{c|}{83\%}         & \multicolumn{1}{c|}{100\%}        & 90\%         & \multicolumn{1}{c|}{91\%}         & \multicolumn{1}{c|}{100\%}        & 95\%         \\ \cline{2-11} 
\multicolumn{1}{|c|}{}                                        & Normal                 & \multicolumn{1}{c|}{100\%}        & \multicolumn{1}{c|}{100\%}        & 100\%        & \multicolumn{1}{c|}{100\%}        & \multicolumn{1}{c|}{100\%}        & 100\%        & \multicolumn{1}{c|}{100\%}        & \multicolumn{1}{c|}{100\%}        & 100\%        \\ \cline{2-11} 
\multicolumn{1}{|c|}{}                                        & Password               & \multicolumn{1}{c|}{69\%}         & \multicolumn{1}{c|}{51\%}         & 44\%         & \multicolumn{1}{c|}{11\%}         & \multicolumn{1}{c|}{44\%}         & 88\%         & \multicolumn{1}{c|}{19\%}         & \multicolumn{1}{c|}{47\%}         & 59\%         \\ \cline{2-11} 
\multicolumn{1}{|c|}{}                                        & Port\_Scanning         & \multicolumn{1}{c|}{0\%}          & \multicolumn{1}{c|}{0\%}          & 0\%          & \multicolumn{1}{c|}{0\%}          & \multicolumn{1}{c|}{0\%}          & 0\%          & \multicolumn{1}{c|}{0\%}          & \multicolumn{1}{c|}{0\%}          & 0\%          \\ \cline{2-11} 
\multicolumn{1}{|c|}{}                                        & Ransomware             & \multicolumn{1}{c|}{0\%}          & \multicolumn{1}{c|}{79\%}         & 0\%          & \multicolumn{1}{c|}{0\%}          & \multicolumn{1}{c|}{24\%}         & 0\%          & \multicolumn{1}{c|}{0\%}          & \multicolumn{1}{c|}{37\%}         & 0\%          \\ \cline{2-11} 
\multicolumn{1}{|c|}{}                                        & SQL\_injection         & \multicolumn{1}{c|}{42\%}         & \multicolumn{1}{c|}{47\%}         & 64\%         & \multicolumn{1}{c|}{90\%}         & \multicolumn{1}{c|}{61\%}         & 17\%         & \multicolumn{1}{c|}{57\%}         & \multicolumn{1}{c|}{53\%}         & 27\%         \\ \cline{2-11} 
\multicolumn{1}{|c|}{}                                        & Uploading              & \multicolumn{1}{c|}{53\%}         & \multicolumn{1}{c|}{65\%}         & 58\%         & \multicolumn{1}{c|}{30\%}         & \multicolumn{1}{c|}{49\%}         & 37\%         & \multicolumn{1}{c|}{38\%}         & \multicolumn{1}{c|}{55\%}         & 45\%         \\ \cline{2-11} 
\multicolumn{1}{|c|}{}                                        & Vulnerability\_scanner & \multicolumn{1}{c|}{93\%}         & \multicolumn{1}{c|}{94\%}         & 94\%         & \multicolumn{1}{c|}{84\%}         & \multicolumn{1}{c|}{85\%}         & 84\%         & \multicolumn{1}{c|}{88\%}         & \multicolumn{1}{c|}{89\%}         & 89\%         \\ \cline{2-11} 
\multicolumn{1}{|c|}{}                                        & XSS                    & \multicolumn{1}{c|}{100\%}        & \multicolumn{1}{c|}{52\%}         & 100\%        & \multicolumn{1}{c|}{2\%}          & \multicolumn{1}{c|}{18\%}         & 3\%          & \multicolumn{1}{c|}{4\%}          & \multicolumn{1}{c|}{27\%}         & 6\%          \\ \hline
\multicolumn{1}{|c|}{\multirow{15}{*}{\begin{tabular}[c]{@{}c@{}}Poisoning \\ Attack \\ $\alpha$ $= 60\%$ \\   \end{tabular}}} 
 & Backdoor               & \multicolumn{1}{c|}{72\%}         & \multicolumn{1}{c|}{0\%}          & 72\%         & \multicolumn{1}{c|}{93\%}         & \multicolumn{1}{c|}{0\%}          & 93\%         & \multicolumn{1}{c|}{82\%}         & \multicolumn{1}{c|}{0\%}          & 81\%         \\ \cline{2-11} 
\multicolumn{1}{|c|}{}                                        & DDoS\_HTTP             & \multicolumn{1}{c|}{70\%}         & \multicolumn{1}{c|}{0\%}          & 73\%         & \multicolumn{1}{c|}{99\%}         & \multicolumn{1}{c|}{0\%}          & 95\%         & \multicolumn{1}{c|}{82\%}         & \multicolumn{1}{c|}{0\%}          & 83\%         \\ \cline{2-11} 
\multicolumn{1}{|c|}{}                                        & DDoS\_ICMP             & \multicolumn{1}{c|}{98\%}         & \multicolumn{1}{c|}{0\%}          & 99\%         & \multicolumn{1}{c|}{97\%}         & \multicolumn{1}{c|}{0\%}          & 100\%        & \multicolumn{1}{c|}{98\%}         & \multicolumn{1}{c|}{0\%}          & 99\%         \\ \cline{2-11} 
\multicolumn{1}{|c|}{}                                        & DDoS\_TCP              & \multicolumn{1}{c|}{69\%}         & \multicolumn{1}{c|}{0\%}          & 68\%         & \multicolumn{1}{c|}{100\%}        & \multicolumn{1}{c|}{0\%}          & 100\%        & \multicolumn{1}{c|}{81\%}         & \multicolumn{1}{c|}{0\%}          & 81\%         \\ \cline{2-11} 
\multicolumn{1}{|c|}{}                                        & DDoS\_UDP              & \multicolumn{1}{c|}{8\%}          & \multicolumn{1}{c|}{8\%}          & 8\%          & \multicolumn{1}{c|}{99\%}         & \multicolumn{1}{c|}{100\%}        & 100\%        & \multicolumn{1}{c|}{15\%}         & \multicolumn{1}{c|}{15\%}         & 15\%         \\ \cline{2-11} 
\multicolumn{1}{|c|}{}                                        & Fingerprinting         & \multicolumn{1}{c|}{0\%}          & \multicolumn{1}{c|}{0\%}          & 0\%          & \multicolumn{1}{c|}{0\%}          & \multicolumn{1}{c|}{0\%}          & 0\%          & \multicolumn{1}{c|}{0\%}          & \multicolumn{1}{c|}{0\%}          & 0\%          \\ \cline{2-11} 
\multicolumn{1}{|c|}{}                                        & MITM                   & \multicolumn{1}{c|}{100\%}        & \multicolumn{1}{c|}{0\%}          & 100\%        & \multicolumn{1}{c|}{88\%}         & \multicolumn{1}{c|}{0\%}          & 90\%         & \multicolumn{1}{c|}{93\%}         & \multicolumn{1}{c|}{0\%}          & 95\%         \\ \cline{2-11} 
\multicolumn{1}{|c|}{}                                        & Normal                 & \multicolumn{1}{c|}{100\%}        & \multicolumn{1}{c|}{0\%}          & 100\%        & \multicolumn{1}{c|}{0\%}          & \multicolumn{1}{c|}{0\%}          & 0\%          & \multicolumn{1}{c|}{0\%}          & \multicolumn{1}{c|}{0\%}          & 0\%          \\ \cline{2-11} 
\multicolumn{1}{|c|}{}                                        & Password               & \multicolumn{1}{c|}{43\%}         & \multicolumn{1}{c|}{0\%}          & 54\%         & \multicolumn{1}{c|}{11\%}         & \multicolumn{1}{c|}{0\%}          & 27\%         & \multicolumn{1}{c|}{58\%}         & \multicolumn{1}{c|}{0\%}          & 36\%         \\ \cline{2-11} 
\multicolumn{1}{|c|}{}                                        & Port\_Scanning         & \multicolumn{1}{c|}{0\%}          & \multicolumn{1}{c|}{0\%}          & 0\%          & \multicolumn{1}{c|}{0\%}          & \multicolumn{1}{c|}{0\%}          & 0\%          & \multicolumn{1}{c|}{0\%}          & \multicolumn{1}{c|}{0\%}          & 0\%          \\ \cline{2-11} 
\multicolumn{1}{|c|}{}                                        & Ransomware             & \multicolumn{1}{c|}{0\%}          & \multicolumn{1}{c|}{0\%}          & 13\%         & \multicolumn{1}{c|}{0\%}          & \multicolumn{1}{c|}{0\%}          & 0\%          & \multicolumn{1}{c|}{0\%}          & \multicolumn{1}{c|}{0\%}          & 0\%          \\ \cline{2-11} 
\multicolumn{1}{|c|}{}                                        & SQL\_injection         & \multicolumn{1}{c|}{57\%}         & \multicolumn{1}{c|}{15\%}         & 43\%         & \multicolumn{1}{c|}{17\%}         & \multicolumn{1}{c|}{61\%}        & 17\%         & \multicolumn{1}{c|}{27\%}         & \multicolumn{1}{c|}{26\%}         & 55\%         \\ \cline{2-11} 
\multicolumn{1}{|c|}{}                                        & Uploading              & \multicolumn{1}{c|}{59\%}         & \multicolumn{1}{c|}{0\%}          & 58\%         & \multicolumn{1}{c|}{30\%}         & \multicolumn{1}{c|}{0\%}          & 37\%         & \multicolumn{1}{c|}{45\%}         & \multicolumn{1}{c|}{0\%}          & 47\%         \\ \cline{2-11} 
\multicolumn{1}{|c|}{}                                        & Vulnerability\_scanner & \multicolumn{1}{c|}{93\%}         & \multicolumn{1}{c|}{0\%}          & 94\%         & \multicolumn{1}{c|}{84\%}         & \multicolumn{1}{c|}{0\%}          & 84\%         & \multicolumn{1}{c|}{88\%}         & \multicolumn{1}{c|}{0\%}          & 89\%         \\ \cline{2-11} 
\multicolumn{1}{|c|}{}                                        & XSS                    & \multicolumn{1}{c|}{0\%}          & \multicolumn{1}{c|}{0\%}          & 53\%         & \multicolumn{1}{c|}{0\%}          & \multicolumn{1}{c|}{0\%}          & 3\%         & \multicolumn{1}{c|}{0\%}          & \multicolumn{1}{c|}{0\%}          & 4\%         \\ \hline
\end{tabular}
\end{table}

\begin{figure}[h!]
\centering
\begin{subfigure}{0.13\textwidth}
  \centering
  \scriptsize
  \includegraphics[width=1\linewidth]{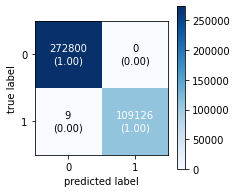}  
  \caption{\scriptsize LR = 0.1 without Poisoning Attack ($\alpha$ $= 0\%$)}
  \label{fig:sub-first}
\end{subfigure}
\hfill
\begin{subfigure}{0.13\textwidth}
  \centering
  \includegraphics[width=1\linewidth]{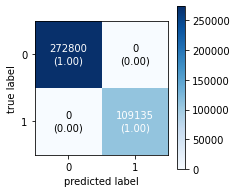}  
  \caption{\scriptsize LR = 0.01 without Poisoning Attack ($\alpha$ $= 0\%$)}
  \label{fig:sub-second}
\end{subfigure}
\hfill
\begin{subfigure}{0.13\textwidth}
  \centering
  \includegraphics[width=1\linewidth]{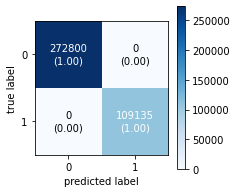}  
  \caption{ \scriptsize LR = 0.001 without Poisoning Attack ($\alpha$ $= 0\%$)}
  \label{fig:sub-second}
\end{subfigure}
\hfill
\begin{subfigure}{0.13\textwidth}
  \centering
  \includegraphics[width=1\linewidth]{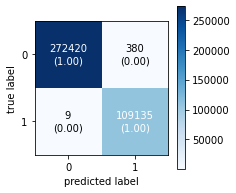}  
  \caption{\scriptsize LR = 0.1 with Poisoning Attack ($\alpha$ $= 60\%$)}
  \label{fig:sub-third}
\end{subfigure}
\hfill
\begin{subfigure}{0.13\textwidth}
  \centering
  \includegraphics[width=1\linewidth]{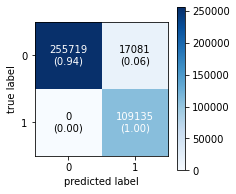}  
  \caption{\scriptsize LR = 0.01 with Poisoning Attack ($\alpha$ $= 60\%$)}
  \label{fig:sub-fourth}
\end{subfigure}
\hfill
\begin{subfigure}{0.13\textwidth}
  \centering
  \includegraphics[width=1\linewidth]{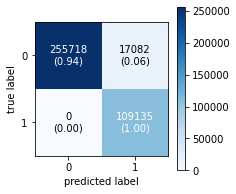}  
  \caption{\scriptsize LR = 0.001 with Poisoning Attack ($\alpha$ $= 60\%$)}
  \label{fig:sub-fourth}
\end{subfigure}
\hfill
\begin{subfigure}{0.13\textwidth}
  \centering
  \includegraphics[width=1\linewidth]{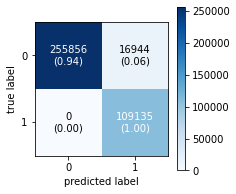}  
  \caption{\scriptsize LR = 0.1 with Poisoning Attack ($\alpha$ $= 80\%$)}
  \label{fig:sub-third}
\end{subfigure}
\hfill
\begin{subfigure}{0.13\textwidth}
  \centering
  \includegraphics[width=1\linewidth]{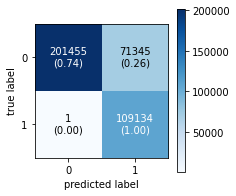}  
  \caption{\scriptsize LR = 0.01 with Poisoning Attack ($\alpha$ $= 80\%$)}
  \label{fig:sub-fourth}
\end{subfigure}
\hfill
\begin{subfigure}{0.13\textwidth}
  \centering
  \includegraphics[width=1\linewidth]{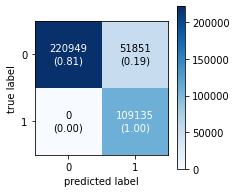}  
  \caption{\scriptsize LR = 0.001 with Poisoning Attack ($\alpha$ $= 80\%$)}
  \label{fig:sub-fourth}
\end{subfigure}

\caption{Confusion matrix of DNN for binary classification in centralized model performance.}
\label{fig:fig10}
\end{figure}

\begin{table*}[h!]

\setlength{\tabcolsep}{2.5pt}
\renewcommand{\arraystretch}{1}
\caption{Accuracy of the federated deep learning approach (CNN) for binary classification and multi-classification in federated model performance with IID data and different numbers of honest clients $K_h$ and malicious clients $K_m$.}
\centering
\label{tab:tab9}
\tiny
\begin{tabular}{|c|cc|cccccccccc|}
\hline
\multirow{3}{*}{\textbf{Classification   mode}} & \multicolumn{2}{c|}{\multirow{2}{*}{\textbf{Clients}}}    & \multicolumn{10}{c|}{\textbf{Federated learning rounds}}                                                                                                                                                                                                                                                                                                                                                                                                                                                                   \\ \cline{4-13} 
                                                & \multicolumn{2}{c|}{}                                     & \multicolumn{1}{c|}{\multirow{2}{*}{\textbf{1}}} & \multicolumn{1}{c|}{\multirow{2}{*}{\textbf{2}}} & \multicolumn{1}{c|}{\multirow{2}{*}{\textbf{3}}} & \multicolumn{1}{c|}{\multirow{2}{*}{\textbf{4}}} & \multicolumn{1}{c|}{\multirow{2}{*}{\textbf{5}}} & \multicolumn{1}{c|}{\multirow{2}{*}{\textbf{6}}} & \multicolumn{1}{c|}{\multirow{2}{*}{\textbf{7}}} & \multicolumn{1}{c|}{\multirow{2}{*}{\textbf{8}}} & \multicolumn{1}{c|}{\multirow{2}{*}{\textbf{9}}} & \multirow{2}{*}{\textbf{10}} \\ \cline{2-3}
                                                & \multicolumn{1}{c|}{\textbf{Honest}} & \textbf{Malicious} & \multicolumn{1}{c|}{}                            & \multicolumn{1}{c|}{}                            & \multicolumn{1}{c|}{}                            & \multicolumn{1}{c|}{}                            & \multicolumn{1}{c|}{}                            & \multicolumn{1}{c|}{}                            & \multicolumn{1}{c|}{}                            & \multicolumn{1}{c|}{}                            & \multicolumn{1}{c|}{}                            &                              \\ \hline
\multirow{3}{*}{Binary   classification}        & \multicolumn{1}{c|}{$K_h=10$}           & $K_m=0$               & \multicolumn{1}{c|}{100\%}                       & \multicolumn{1}{c|}{100\%}                       & \multicolumn{1}{c|}{100\%}                       & \multicolumn{1}{c|}{100\%}                       & \multicolumn{1}{c|}{100\%}                       & \multicolumn{1}{c|}{100\%}                       & \multicolumn{1}{c|}{100\%}                       & \multicolumn{1}{c|}{100\%}                       & \multicolumn{1}{c|}{100\%}                       & 100\%                        \\ \cline{2-13} 
                                                & \multicolumn{1}{c|}{$K_h=7$}         & $K_m=3$               & \multicolumn{1}{c|}{100\%}                       & \multicolumn{1}{c|}{100\%}                       & \multicolumn{1}{c|}{100\%}                       & \multicolumn{1}{c|}{100\%}                       & \multicolumn{1}{c|}{100\%}                       & \multicolumn{1}{c|}{100\%}                       & \multicolumn{1}{c|}{100\%}                       & \multicolumn{1}{c|}{100\%}                       & \multicolumn{1}{c|}{100\%}                       & 100\%                        \\ \cline{2-13} 
                                                & \multicolumn{1}{c|}{$K_h=3$}         & $K_m=7$               & \multicolumn{1}{c|}{80.41\%}                     & \multicolumn{1}{c|}{80.52\%}                     & \multicolumn{1}{c|}{80.52\%}                     & \multicolumn{1}{c|}{96.25\%}                     & \multicolumn{1}{c|}{87.90\%}                     & \multicolumn{1}{c|}{87.90\%}                     & \multicolumn{1}{c|}{100\%}                       & \multicolumn{1}{c|}{95.59\%}                     & \multicolumn{1}{c|}{100\%}                       & 100\%                        \\ \hline
\multirow{3}{*}{Multi   classification}         & \multicolumn{1}{c|}{$K_h=10$}           & $K_m=0$               & \multicolumn{1}{c|}{82.44\%}                     & \multicolumn{1}{c|}{93.80\%}                     & \multicolumn{1}{c|}{94.04\%}                     & \multicolumn{1}{c|}{94.11\%}                     & \multicolumn{1}{c|}{94.14\%}                     & \multicolumn{1}{c|}{94.25\%}                     & \multicolumn{1}{c|}{94.64\%}                     & \multicolumn{1}{c|}{94.73\%}                     & \multicolumn{1}{c|}{94.91\%}                     & 94.93\%                      \\ \cline{2-13} 
                                                & \multicolumn{1}{c|}{$K_h=7$}         & $K_m=3$               & \multicolumn{1}{c|}{76.86\%}                     & \multicolumn{1}{c|}{89.04\%}                     & \multicolumn{1}{c|}{89.52\%}                     & \multicolumn{1}{c|}{92.48\%}                     & \multicolumn{1}{c|}{17.09\%}                     & \multicolumn{1}{c|}{94.41\%}                     & \multicolumn{1}{c|}{80.87\%}                     & \multicolumn{1}{c|}{92.79\%}                     & \multicolumn{1}{c|}{20.91\%}                     & 94.33\%                      \\ \cline{2-13} 
                                                & \multicolumn{1}{c|}{$K_h=3$}         & $K_m=7$               & \multicolumn{1}{c|}{4.99\%}                      & \multicolumn{1}{c|}{5.00\%}                      & \multicolumn{1}{c|}{15.94\%}                     & \multicolumn{1}{c|}{86.47\%}                     & \multicolumn{1}{c|}{6.73\%}                      & \multicolumn{1}{c|}{86.14\%}                     & \multicolumn{1}{c|}{5.14\%}                      & \multicolumn{1}{c|}{85.85\%}                     & \multicolumn{1}{c|}{7.55\%}                      & 85.98\%                      \\ \hline
                            \end{tabular}
\end{table*}

\begin{table*}[h!]

\setlength{\tabcolsep}{2.5pt}
\renewcommand{\arraystretch}{1}
\caption{Accuracy of the federated deep learning approach (CNN) for binary classification and multi-classification in federated model performance with Non-IID data.}
\centering
\label{tab:tab10}
\tiny
\begin{tabular}{|c|cc|cccccccccc|}
\hline
\multirow{3}{*}{\textbf{Classification mode}} & \multicolumn{2}{c|}{\multirow{2}{*}{\textbf{Clients}}}    & \multicolumn{10}{c|}{\textbf{Federated learning rounds}}                                                                                                                                                                                                                                                                                                                                                                                                                                                                   \\ \cline{4-13} 
                                              & \multicolumn{2}{c|}{}                                     & \multicolumn{1}{c|}{\multirow{2}{*}{\textbf{1}}} & \multicolumn{1}{c|}{\multirow{2}{*}{\textbf{2}}} & \multicolumn{1}{c|}{\multirow{2}{*}{\textbf{3}}} & \multicolumn{1}{c|}{\multirow{2}{*}{\textbf{4}}} & \multicolumn{1}{c|}{\multirow{2}{*}{\textbf{5}}} & \multicolumn{1}{c|}{\multirow{2}{*}{\textbf{6}}} & \multicolumn{1}{c|}{\multirow{2}{*}{\textbf{7}}} & \multicolumn{1}{c|}{\multirow{2}{*}{\textbf{8}}} & \multicolumn{1}{c|}{\multirow{2}{*}{\textbf{9}}} & \multirow{2}{*}{\textbf{10}} \\ \cline{2-3}
                                              & \multicolumn{1}{c|}{\textbf{Honest}} & \textbf{Malicious} & \multicolumn{1}{c|}{}                            & \multicolumn{1}{c|}{}                            & \multicolumn{1}{c|}{}                            & \multicolumn{1}{c|}{}                            & \multicolumn{1}{c|}{}                            & \multicolumn{1}{c|}{}                            & \multicolumn{1}{c|}{}                            & \multicolumn{1}{c|}{}                            & \multicolumn{1}{c|}{}                            &                              \\ \hline
\multirow{3}{*}{Binary classification}        & \multicolumn{1}{c|}{$K_h=10$}           & $K_m=0$               & \multicolumn{1}{c|}{100\%}                       & \multicolumn{1}{c|}{100\%}                       & \multicolumn{1}{c|}{100\%}                       & \multicolumn{1}{c|}{100\%}                       & \multicolumn{1}{c|}{100\%}                       & \multicolumn{1}{c|}{100\%}                       & \multicolumn{1}{c|}{100\%}                       & \multicolumn{1}{c|}{100\%}                       & \multicolumn{1}{c|}{100\%}                       & 100\%                        \\ \cline{2-13} 
                                              & \multicolumn{1}{c|}{$K_h=7$}           & $K_m=3$               & \multicolumn{1}{c|}{100\%}                       & \multicolumn{1}{c|}{100\%}                       & \multicolumn{1}{c|}{89.29\%}                     & \multicolumn{1}{c|}{98.08\%}                     & \multicolumn{1}{c|}{98.08\%}                     & \multicolumn{1}{c|}{98.08\%}                     & \multicolumn{1}{c|}{98.09\%}                     & \multicolumn{1}{c|}{98.37\%}                     & \multicolumn{1}{c|}{100\%}                       & 100\%                        \\ \cline{2-13} 
                                              & \multicolumn{1}{c|}{$K_h=3$}           & $K_m=7$               & \multicolumn{1}{c|}{26.79\%}                     & \multicolumn{1}{c|}{79.66\%}                     & \multicolumn{1}{c|}{26.79\%}                     & \multicolumn{1}{c|}{99.99\%}                     & \multicolumn{1}{c|}{99.99\%}                     & \multicolumn{1}{c|}{100.00\%}                    & \multicolumn{1}{c|}{100.00\%}                    & \multicolumn{1}{c|}{100.00\%}                    & \multicolumn{1}{c|}{100.00\%}                    & 100.00\%                     \\ \hline
\multirow{3}{*}{Multi classification}         & \multicolumn{1}{c|}{$K_h=10$}           & $K_m=0$               & \multicolumn{1}{c|}{92.28\%}                     & \multicolumn{1}{c|}{93.96\%}                     & \multicolumn{1}{c|}{94.05\%}                     & \multicolumn{1}{c|}{94.06\%}                     & \multicolumn{1}{c|}{94.06\%}                     & \multicolumn{1}{c|}{94.12\%}                     & \multicolumn{1}{c|}{94.12\%}                     & \multicolumn{1}{c|}{94.16\%}                     & \multicolumn{1}{c|}{94.17\%}                     & 94.18\%                      \\ \cline{2-13} 
                                              & \multicolumn{1}{c|}{$K_h=7$}           & $K_m=3$               & \multicolumn{1}{c|}{84.77\%}                     & \multicolumn{1}{c|}{93.18\%}                     & \multicolumn{1}{c|}{72.66\%}                     & \multicolumn{1}{c|}{93.94}                       & \multicolumn{1}{c|}{93.52\%}                     & \multicolumn{1}{c|}{94.11\%}                     & \multicolumn{1}{c|}{93.34\%}                     & \multicolumn{1}{c|}{94.04\%}                     & \multicolumn{1}{c|}{93.70\%}                     & 94.00\%                      \\ \cline{2-13} 
                                              & \multicolumn{1}{c|}{$K_h=3$}           & $K_m=7$               & \multicolumn{1}{c|}{86.64\%}                     & \multicolumn{1}{c|}{13.13\%}                     & \multicolumn{1}{c|}{91.31\%}                     & \multicolumn{1}{c|}{12.04\%}                     & \multicolumn{1}{c|}{92.63\%}                     & \multicolumn{1}{c|}{17.20\%}                     & \multicolumn{1}{c|}{93.53\%}                     & \multicolumn{1}{c|}{75.73\%}                     & \multicolumn{1}{c|}{93.99\%}                     & 30.04\%                      \\ \hline
\end{tabular}

\end{table*}

\begin{table}[h!]

\setlength{\tabcolsep}{2.5pt}
\renewcommand{\arraystretch}{1}
\caption{Federated deep learning approach (CNN) evaluation results for multi-classification with IID and Non-IID data.}
\centering
\label{tab:tab11}
\tiny
\begin{tabular}{|c|c|ccc|ccc|}
\hline
\multirow{2}{*}{\textbf{The attack rate}}                                                                                                               & \multirow{2}{*}{\textbf{Class}} & \multicolumn{3}{c|}{\textbf{IID}}                                                                  & \multicolumn{3}{c|}{\textbf{Non-IID}}                                                              \\ \cline{3-8} 
                                                                                                                                                        &                                 & \multicolumn{1}{c|}{\textbf{Precision}} & \multicolumn{1}{c|}{\textbf{Recall}} & \textbf{F1-score} & \multicolumn{1}{c|}{\textbf{Precision}} & \multicolumn{1}{c|}{\textbf{Recall}} & \textbf{F1-score} \\ \hline
\multirow{15}{*}{\begin{tabular}[c]{@{}c@{}}No attack\\    \\  \\    \\ $\alpha$ $= 0\%$\\    \\  \\    \\ $K_h=10$              $K_m=0$\end{tabular}}               & Backdoor                        & \multicolumn{1}{c|}{82\%}               & \multicolumn{1}{c|}{95\%}            & 88\%              & \multicolumn{1}{c|}{73\%}               & \multicolumn{1}{c|}{94\%}            & 82\%              \\ \cline{2-8} 
                                                                                                                                                        & DDoS\_HTTP                      & \multicolumn{1}{c|}{74\%}               & \multicolumn{1}{c|}{95\%}            & 83\%              & \multicolumn{1}{c|}{73\%}               & \multicolumn{1}{c|}{96\%}            & 83\%              \\ \cline{2-8} 
                                                                                                                                                        & DDoS\_ICMP                      & \multicolumn{1}{c|}{100\%}              & \multicolumn{1}{c|}{100\%}           & 100\%             & \multicolumn{1}{c|}{99\%}               & \multicolumn{1}{c|}{100\%}           & 100\%             \\ \cline{2-8} 
                                                                                                                                                        & DDoS\_TCP                       & \multicolumn{1}{c|}{82\%}               & \multicolumn{1}{c|}{100\%}           & 90\%              & \multicolumn{1}{c|}{70\%}               & \multicolumn{1}{c|}{100\%}           & 83\%              \\ \cline{2-8} 
                                                                                                                                                        & DDoS\_UDP                       & \multicolumn{1}{c|}{100\%}              & \multicolumn{1}{c|}{100\%}           & 100\%             & \multicolumn{1}{c|}{100\%}              & \multicolumn{1}{c|}{100\%}           & 100\%             \\ \cline{2-8} 
                                                                                                                                                        & Fingerprinting                  & \multicolumn{1}{c|}{0\%}                & \multicolumn{1}{c|}{0\%}             & 0\%               & \multicolumn{1}{c|}{0\%}                & \multicolumn{1}{c|}{0\%}             & 0\%               \\ \cline{2-8} 
                                                                                                                                                        & MITM                            & \multicolumn{1}{c|}{100\%}              & \multicolumn{1}{c|}{97\%}            & 99\%              & \multicolumn{1}{c|}{100\%}              & \multicolumn{1}{c|}{100\%}           & 100\%             \\ \cline{2-8} 
                                                                                                                                                        & Normal                          & \multicolumn{1}{c|}{100\%}              & \multicolumn{1}{c|}{100\%}           & 100\%             & \multicolumn{1}{c|}{100\%}              & \multicolumn{1}{c|}{100\%}           & 100\%             \\ \cline{2-8} 
                                                                                                                                                        & Password                        & \multicolumn{1}{c|}{57\%}               & \multicolumn{1}{c|}{31\%}            & 40\%              & \multicolumn{1}{c|}{43\%}               & \multicolumn{1}{c|}{84\%}            & 57\%              \\ \cline{2-8} 
                                                                                                                                                        & Port\_Scanning                  & \multicolumn{1}{c|}{92\%}               & \multicolumn{1}{c|}{52\%}            & 66\%              & \multicolumn{1}{c|}{67\%}               & \multicolumn{1}{c|}{8\%}             & 14\%              \\ \cline{2-8} 
                                                                                                                                                        & Ransomware                      & \multicolumn{1}{c|}{77\%}               & \multicolumn{1}{c|}{47\%}            & 58\%              & \multicolumn{1}{c|}{53\%}               & \multicolumn{1}{c|}{16\%}            & 25\%              \\ \cline{2-8} 
                                                                                                                                                        & SQL\_injection                  & \multicolumn{1}{c|}{44\%}               & \multicolumn{1}{c|}{77\%}            & 56\%              & \multicolumn{1}{c|}{56\%}               & \multicolumn{1}{c|}{17\%}            & 26\%              \\ \cline{2-8} 
                                                                                                                                                        & Uploading                       & \multicolumn{1}{c|}{62\%}               & \multicolumn{1}{c|}{38\%}            & 47\%              & \multicolumn{1}{c|}{60\%}               & \multicolumn{1}{c|}{39\%}            & 48\%              \\ \cline{2-8} 
                                                                                                                                                        & Vulnerability\_scanner          & \multicolumn{1}{c|}{94\%}               & \multicolumn{1}{c|}{84\%}            & 89\%              & \multicolumn{1}{c|}{94\%}               & \multicolumn{1}{c|}{84\%}            & 89\%              \\ \cline{2-8} 
                                                                                                                                                        & XSS                             & \multicolumn{1}{c|}{56\%}               & \multicolumn{1}{c|}{20\%}            & 29\%              & \multicolumn{1}{c|}{56\%}               & \multicolumn{1}{c|}{20\%}            & 29\%              \\ \hline
\multirow{15}{*}{\begin{tabular}[c]{@{}c@{}}Poisoning attack\\    \\  \\    \\ $\alpha$ $= 60\%$\\    \\  \\    \\ $K_h=3$              $K_m=7$\end{tabular}} & Backdoor                        & \multicolumn{1}{c|}{3\%}                & \multicolumn{1}{c|}{1\%}             & 2\%               & \multicolumn{1}{c|}{0\%}                & \multicolumn{1}{c|}{0\%}             & 0\%               \\ \cline{2-8} 
                                                                                                                                                        & DDoS\_HTTP                      & \multicolumn{1}{c|}{69\%}               & \multicolumn{1}{c|}{95\%}            & 80\%              & \multicolumn{1}{c|}{42\%}               & \multicolumn{1}{c|}{96\%}            & 59\%              \\ \cline{2-8} 
                                                                                                                                                        & DDoS\_ICMP                      & \multicolumn{1}{c|}{0\%}                & \multicolumn{1}{c|}{0\%}             & 0\%               & \multicolumn{1}{c|}{100\%}              & \multicolumn{1}{c|}{39\%}            & 56\%              \\ \cline{2-8} 
                                                                                                                                                        & DDoS\_TCP                       & \multicolumn{1}{c|}{75\%}               & \multicolumn{1}{c|}{52\%}            & 61\%              & \multicolumn{1}{c|}{74\%}               & \multicolumn{1}{c|}{29\%}            & 41\%              \\ \cline{2-8} 
                                                                                                                                                        & DDoS\_UDP                       & \multicolumn{1}{c|}{37\%}               & \multicolumn{1}{c|}{100\%}           & 54\%              & \multicolumn{1}{c|}{7\%}                & \multicolumn{1}{c|}{100\%}           & 13\%              \\ \cline{2-8} 
                                                                                                                                                        & Fingerprinting                  & \multicolumn{1}{c|}{0\%}                & \multicolumn{1}{c|}{0\%}             & 0\%               & \multicolumn{1}{c|}{0\%}                & \multicolumn{1}{c|}{0\%}             & 0\%               \\ \cline{2-8} 
                                                                                                                                                        & MITM                            & \multicolumn{1}{c|}{100\%}              & \multicolumn{1}{c|}{93\%}            & 96\%              & \multicolumn{1}{c|}{0\%}                & \multicolumn{1}{c|}{0\%}             & 0\%               \\ \cline{2-8} 
                                                                                                                                                        & Normal                          & \multicolumn{1}{c|}{100\%}              & \multicolumn{1}{c|}{100\%}           & 100\%             & \multicolumn{1}{c|}{50\%}              & \multicolumn{1}{c|}{23\%}            & 38\%              \\ \cline{2-8} 
                                                                                                                                                        & Password                        & \multicolumn{1}{c|}{40\%}               & \multicolumn{1}{c|}{30\%}            & 48\%              & \multicolumn{1}{c|}{41\%}               & \multicolumn{1}{c|}{33\%}            & 36\%              \\ \cline{2-8} 
                                                                                                                                                        & Port\_Scanning                  & \multicolumn{1}{c|}{31\%}               & \multicolumn{1}{c|}{4\%}             & 7\%               & \multicolumn{1}{c|}{0\%}                & \multicolumn{1}{c|}{0\%}             & 0\%               \\ \cline{2-8} 
                                                                                                                                                        & Ransomware                      & \multicolumn{1}{c|}{0\%}                & \multicolumn{1}{c|}{0\%}             & 0\%               & \multicolumn{1}{c|}{0\%}                & \multicolumn{1}{c|}{0\%}             & 0\%               \\ \cline{2-8} 
                                                                                                                                                        & SQL\_injection                  & \multicolumn{1}{c|}{44\%}               & \multicolumn{1}{c|}{29\%}            & 35\%              & \multicolumn{1}{c|}{38\%}               & \multicolumn{1}{c|}{17\%}            & 24\%              \\ \cline{2-8} 
                                                                                                                                                        & Uploading                       & \multicolumn{1}{c|}{71\%}               & \multicolumn{1}{c|}{21\%}            & 32\%              & \multicolumn{1}{c|}{0\%}                & \multicolumn{1}{c|}{0\%}             & 0\%               \\ \cline{2-8} 
                                                                                                                                                        & Vulnerability\_scanner          & \multicolumn{1}{c|}{76\%}               & \multicolumn{1}{c|}{86\%}            & 81\%              & \multicolumn{1}{c|}{80\%}               & \multicolumn{1}{c|}{84\%}            & 82\%              \\ \cline{2-8} 
                                                                                                                                                        & XSS                             & \multicolumn{1}{c|}{64\%}               & \multicolumn{1}{c|}{7\%}             & 12\%              & \multicolumn{1}{c|}{92\%}               & \multicolumn{1}{c|}{4\%}             & 8\%               \\ \hline
\end{tabular}
\end{table}

\section{Experimental Evaluation}
\label{sec:6}

\subsection{Experimental setup}

We conducted an experimental analysis of poisoning attacks against intrusion detection systems based on deep learning approaches in centralized and federated learning. We choose three deep learning approaches: DNN, CNN, and RNN. The classification tasks are conducted in two modes: Binary classification and Multi-class classification. The Binary classification includes two classes (i.e., Normal or Attack). The Multi-class classification includes 15 classes (i.e., Normal or attack types). We use Google Colab with Python libraries to analyze and visualize data. To build, train, and evaluate poisoning models, we use both open-source frameworks, namely, Keras and PyTorch. We adopt both the IID and Non-IID data distribution in federated learning.  Table~\ref{tab:tab12} presents the details of settings for experimental evaluation.

\subsection{Dataset description and pre-processing}

We use the Edge-IIoTset dataset \cite{ferrag2022edge}, a new comprehensive, realistic cyber security dataset. The Edge-IIoTset dataset is generated using a purpose-built IoT/IIoT testbed. The testbed consists of seven interconnected layers: IoT/IIoT perception layer, edge layer, SDN layer, fog layer, Blockchain layer, NFV layer, and cloud computing layer. It contains more than 20 million total instances for normal and attacks traffic in CSV and PCAP files, with over 63 features. The pre-processing data phases are organized into the following seven steps: 

\begin{enumerate}
    \item Clean corrupted and duplicated rows.
    \item Clean unnecessary columns (features), especially for avoiding overfitting.
    \item Encode categorical features as a one-hot numeric array using the $OneHotEncoder()$ function.
    \item Split the dataset into random train (80\%) and test (20\%) subsets. 
    \item Standardize features using the $StandardScaler()$ function.
    \item Perform oversampling using SMOTE \footnote{To oversample data in minority classes while avoiding overfitting, we use the Synthetic Minority Over-sampling Technique (SMOTE). }.
    \item Give a new shape to an array without changing its data, which is used for RNN and CNN\footnote{The reason for reshaping is to provide the correct data to the CNN and RNN models using the numpy.reshape() function. }.
\end{enumerate}


\subsection{Performance Metrics}

In order to evaluate the performance of machine learning models, we use the following important performance metrics:

\begin{itemize}
\item \textit{True Positive (TP)}: correctly classified attack samples.
    \item \textit{False Negative (FN)}: wrongly classified attack samples.
    \item \textit{True Negative (TN)}: correctly classified benign samples.
    \item \textit{False Positive (FP)}: wrongly classified benign samples.
    \item \textit{Accuracy}, given by: $ \frac{TP_{Attack}+TN_{Normal}}{TP_{Attack}+TN_{Normal}+FP_{Normal}+FN_{Attack}}$
    \item \textit{Precision}, given by: $\frac{TP_{Attack}}{TP_{Attack} + FP_{Normal}}$
    \item \textit{Recall}, given by: $ \frac{TP_{Attack}}{TP_{Attack} + FN_{Attack}}$
    \item \textit{$F_{1}$-Score}, given by: $2 \cdot \frac{Precision \cdot Recall}{Precision+Recall}$
    \item \textit{Poisoning attack rate}: is used to measure the success of the poisoning attack for each label, which is given by: $1-\frac{Recall_{wp}}{Recall_{wop}}$

   
\end{itemize}

 \noindent where $Recall_{wp}$ is the detection rate of the intrusion attack after the poisoning attack and $Recall_{wop}$ is the detection rate of the intrusion attack before the poisoning attack.

\begin{figure}[h!]
\centering
\centering
     \begin{subfigure}[b]{0.45\textwidth}
         \centering
\begin{tikzpicture}[scale=0.65,line width=1pt]
\begin{axis}[xbar = .001cm,
x tick label style={
		/pgf/number format/1000 sep=},
    xlabel=Poisoning attack rate(\%),
    bar width = 3pt,
    xmin = 0,
    xmax = 90,
    ymin=0, ymax=15,
    ytick = data,
    yticklabels={Backdoor,DDoS\_HTTP,DDoS\_ICMP,DDoS\_TCP, DDoS\_UDP,Fingerprinting,MITM,Normal,Password, Port\_Scanning,Ransomware,SQL\_injection,Uploading,Vulnerability\_scanner,XSS},
    tick label style={font=\scriptsize},
    enlarge x limits = {value = .25, upper},
    enlarge y limits = {abs = .8},
    legend style={at={(0.5,-0.2)},
		anchor=north,legend columns=-1}
]
 
\addplot coordinates {(98.94,1) (0,2) (100,3) (48,4) (0,5) (0,6) (1.03,7) (50,8) (3.22,9)(86.53,10)(100,11)(54,12)(15.78,13) (3.57,14)(40,15)};
\addplot coordinates {(100,1) (0,2) (61,3) (71,4) (0,5) (0,6) (100,7) (77,8) (60.71,9)(100,10)(100,11)(0,12)(100,13)(0,14)(80,15)};

\legend {CNN IID, CNN Non-IID};

\end{axis}

\end{tikzpicture}
\label{fig:y equals x}
\caption{Federated Edge Learning}
     \end{subfigure}
   
\centering
     \begin{subfigure}[b]{0.45\textwidth}
         \centering
        \begin{tikzpicture}[scale=0.65,line width=1pt] 
\begin{axis}[xbar = 0.01pt,
enlargelimits=0.5,
x tick label style={
		/pgf/number format/1000 sep=},
    xlabel=Poisoning attack rate(\%),
    bar width = 3pt,
    xmin = 0,
    xmax = 90,
    ymin=0, ymax=15,
    ytick = data,
    tick label style={font=\scriptsize},
    yticklabels={Backdoor,DDoS\_HTTP,DDoS\_ICMP,DDoS\_TCP, DDoS\_UDP,Fingerprinting,MITM,Normal,Password, Port\_Scanning,Ransomware,SQL\_injection,Uploading,Vulnerability\_scanner,XSS},
    enlarge x limits = {value = .30, upper},
    enlarge y limits = {abs = .20},
    legend style={at={(0.5,-0.2)},
		anchor=north,legend columns=-1}
]
 
\addplot[xbar,fill=blue]  coordinates {(0,1) (1.01,2) (1.02,3) (0,4) (0,5) (0,6) (6.02,7) (100,8) (0,9)(0,10)(0,11)(81.11,12)(0,13) (0,14)(100,15)};
\addplot[xbar,fill=green] coordinates {(100,1) (100,2) (100,3) (0,4) (100,5) (100,6) (100,7) (100,8) (0,9)(0,10)(100,11)(0,12)(100,13) (100,14)(100,15)};
\addplot[xbar,fill=red] coordinates {(0,1) (4.04,2) (0,3) (0,4) (0,5) (0,6) (0,7) (100,8) (69.31,9)(0,10)(0,11)(0,12)(0,13)(0,14)(0,15)};

\legend {DNN, RNN, CNN};
 
\end{axis}

\end{tikzpicture}
\label{fig:y equals x}
\caption{Centralized Learning}

     \end{subfigure}
\caption{Evaluation of poisoning attack rate (\%).}
\label{fig:fig2}
\end{figure}
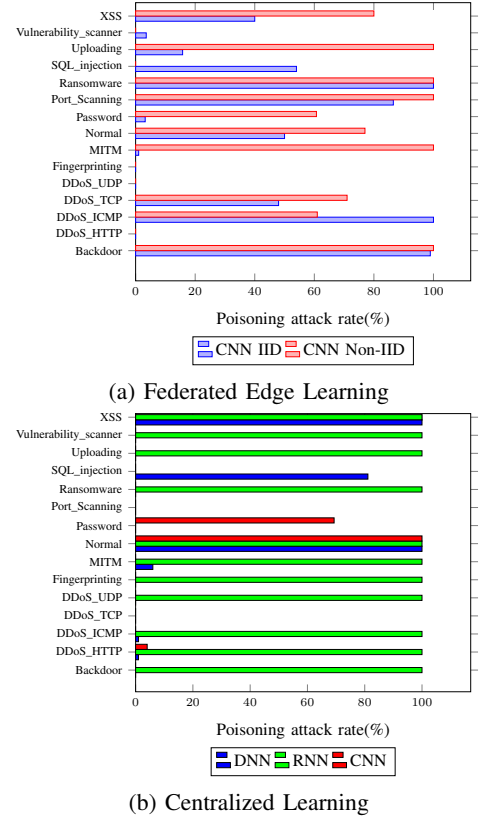

\subsection{Experimental Results}

\subsubsection{Centralized model performance}

Table~\ref{tab:tab9} presents the classification report of the accuracy of deep learning for binary classification and multi-classification under different deep learning approaches, namely, DNN, RNN, and CNN in centralized model performance with the attack rate $\alpha$ $= [0\%, 40\%, 50\%, 60\%]$ (i.e., Poisoning Attack). With binary classification, the accuracy of the DNN classifier is decreased from 100\% to 95.50\%, while with multi-classification, the accuracy is decreased from 93.01\% to 21.76\%. We observe that intrusion detection models based on the deep learning classifiers (i.e., CNN, RNN, DNN) are more affected by multi-classification than binary classification.

Table \ref{tab:tab8} presents the classification report for a multi-class of different deep learning approaches, namely, DNN, RNN, and CNN in centralized model performance with the attack rate $\alpha$ $= 60\%$ (i.e., Poisoning Attack). We observe that the DNN gives the highest precision rate without poisoning attack (i.e., $\alpha$ $= 0\%$) for Normal traffic and three types of attacks, namely, MITM attack 100\%, Password attack 69\%, and XSS attack 100\%. In addition, we observe that the CNN gives the highest precision rate for Normal traffic and three types of attacks: SQL injection attack 64\%, Vulnerability scanner attack 94\%, and XSS attack 100\%.

Figure~\ref{fig:fig10} illustrates the confusion matrix of DNN for binary classification in centralized model performance without poisoning attack $\alpha$ $= 0\%$, with Poisoning Attack $\alpha$ $= 60\%$ and $\alpha$ $= 80\%$, and learning rate LR = [0.1, 0.01, 0.001]. Without poisoning attack (i.e., $\alpha$ $= 0\%$), we observe satisfactory results under the three values of the learning rate LR = [0.1, 0.01, 0.001]. When attackers launch a Poisoning Attack with $\alpha$ $= 60\%$ and $\alpha$ $= 80\%$, we observe that all the DNN classifier is affected and give negative results under the three values of the learning rate LR = [0.1, 0.01, 0.001]. In addition, we observe that when the learning rate is fixed at 0.1, the DNN classifier can resist the Poisoning Attack compared to the results with the learning rate is 0.01 and 0.001.

\subsubsection{Federated model performance}

Table \ref{tab:tab9} presents the accuracy results of the federated deep learning approach (CNN) for binary classification and multi-classification in federated model performance with the IID data and different numbers of honest clients $K_h$ and malicious clients $K_m$. When the number of honest clients is higher than the number of malicious clients (i.e., [$K_h=10$ and $K_m=0$], [$K_h=7$ and $K_m=3$]), we can observe that in each round, the accuracy results increases until it reaches 94.93\% and 100\% with multi-classification and binary classification, respectively. Therefore, when the number of honest clients is less than the number of malicious clients (i.e., $K_h=3$ and $K_m=7$), we can observe that the accuracy results are affected in each round and give negative results until it reaches 85.98\% with multi-classification. 

Table \ref{tab:tab10} presents the accuracy results of the federated deep learning approach (CNN) for binary classification and multi-classification in federated model performance with the Non-IID data and different numbers of honest clients $K_h$ and malicious clients $K_m$. When the number of honest clients is less than the number of malicious clients (i.e., $K_h=3$ and $K_m=7$), we can observe that in each round, the accuracy results are affected in each round and give negative results until it reaches 30.04\%.

Table \ref{tab:tab11} presents the evaluation results of the federated deep learning approach (CNN) for multi-classification in federated model performance with the IID and Non-IID data and different numbers of honest clients $K_h$ and malicious clients $K_m$.  with a poisoning attack (i.e., $\alpha$ $= 60\%$) and the number of honest clients is less than the number of malicious clients (i.e., $K_h=3$ and $K_m=7$), we observe that the CNN classifier is affected and give negative results with the three performance metrics, namely, precision, recall, and F1 score.

Figure \ref{fig:fig2} presents the evaluation results of poisoning attack rate(\%) in two different learning modes, namely, (a) federated model performance under two data distribution types, namely, IID and Non-IID data; (b) centralized model performance with DNN, RNN, and CNN models. The numbers of honest clients $K_h=3$ and malicious clients $K_m=7$. With the federated edge learning setting, we observe the success of the poisoning attack up to 100\% for some classes, which means that the malicious IoT devices have disturbed the privacy-preserving federated learning in both IID and Non-IID modes. The results we observe for the centralized learning settings are, as we would expect, the success of the poisoning attack up to 100\% for the Normal class.

\section{Conclusions} \label{sec:8}

We have proposed an anticipatory study for poisoning attacks in federated edge learning for digital twin 6G-enabled IoTs. We examined the influence of adversaries on the training and development of federated learning models for digital twin 6G-enabled IoTs. We demonstrated that attackers who conduct poisoning attacks in two different learning modes, namely, centralized learning and federated learning, can severely reduce the model's accuracy. We comprehensively evaluated our attacks on a new cyber security dataset designed for IoT applications. The study demonstrates that attackers who conduct poisoning attacks can lead to a decrease in accuracy from 94.93\% to 85.98\% with IID data and from 94.18\% to 30.04\% with Non-IID data.


Since we envision the possibility that attackers will use generative AI to create adversarial samples, our ongoing research agenda includes building efficient approaches to defend against such attacks. Additionally, the protection of the integrity of the data as well as the integrity of the AI model in the federated edge learning models for digital twin 6G-enabled IoTs is also on our radar.




\bibliographystyle{IEEEtran}
\bibliography{ref} 

\end{document}